# Basis-Independent Geometric Phase Modulation in Circularly Birefringent Plasmonic Structures

Pasha Goz [1] Akash Das [1,2] Andre Yaroshevsky [2] and Yuri Gorodetski* [1,2]

[1] Department of Mechatronics and Mechanical Engineering, Ariel University, Ariel 4070000, Israel

[2] Department of Electrical and Electronics Engineering, Ariel University, Ariel 4070000, Israel

*Email: yurig@ariel.ac.il

## Abstract

Geometric phase in metasurfaces is conventionally realized using spatially rotated linearly birefringent meta-atoms under circularly polarized illumination. Here we demonstrate that full $\sim 2\pi$ geometric phase modulation can be achieved through circular birefringence and observed under linear polarization excitation. We introduce a plasmonic metasurface composed of spatially rotated chiral spiral unit cells designed to produce space-variant circular retardance. This generates a geometric phase ramp equivalent to a blazed grating, leading to $\pm$ 1 diffraction orders in momentum space. Using leakage radiation microscopy, we directly resolve these orders and show how their intensities depend upon the input linear polarization confirming the geometric origin of the phase. We theoretically analyze the phenomenon using a rotated Poincaré space and confirm our results by direct Stokes parameters measurement. These results establish basis-independent approach to geometric phase accumulation in circularly birefringent plasmonic metasurfaces.

# Keywords

Geometric phase; circular birefringence; chiral metasurfaces; plasmonics; leakage radiation microscopy; Poincaré sphere

# Introducion

Geometric phase, originating from the seminal works of Pancharatnam and Berry [1, 2] constitutes one of the most profound manifestations of gauge structure in wave physics. In optics, it arises when the polarization state undergoes cyclic evolution on the Poincaré sphere, leading to a phase shift determined solely by the geometry (enclosed solid angle) rather than optical path length [11—5]. This Pancharatnam-Berry (PB) phase has evolved from a conceptual curiosity into a central design principle for flat optics, singular optics, and metasurfaces [6–11].

Conventional PB metasurfaces rely on spatially rotated linearly birefringent meta-atoms [16, 12—14]. Under circularly polarized illumination, the local optical axis orientation $\theta(x)$ imparts a helicity-dependent phase shift $\pm 2\theta$, enabling wavefront shaping via geometric phase gradients [17, 9, 10, 15]. In this widely adopted framework, geometric phase is typically described in the circular polarization basis and is intrinsically associated with spin conversion. However, geometric phase is fundamentally basis-independent: it depends only on the trajectory of the polarization state on the Poincaré sphere [12, 5, 16—18]. This raises a critical question that remains insufficiently explored [19]: can full $2\pi$ geometric phase modulation be achieved through circular birefringence instead of the conventional linear birefringence? Establishing such complementarity would extend geometric-phase optics beyond the conventional PB paradigm and clarify the underlying gauge structure governing metasurface phase engineering.

Chiral metastructurcs provide a natural platform for addressing this question [19, 20]. Subwavelength spiral and twisted meta-atoms can exhibit strong circular birefringence and optical activity [21—23], with circular polarization eigenmodes as the natural basis. However, no direct experimental demonstration of spatially varying geometric phase induced by circular birefringence and its quantitative verification via solid angle reconstruction has been reported. In such systems, spatial variation of the structural orientation induces a position-dependent geometric phase. When implemented in plasmonic platforms, the resulting phase gradient modifies the in-plane momentum of excited surface plasmon (SP) polaritons [24, 25], leading to controlled diffraction into discrete k-space channels [19]. Because SPs exhibit transverse spin and spin momentum locking [26—28], they provide a sensitive probe of polarization-dependent geometric effects at subwavelength scales.

Here we introduce a plasmonic metasurface composed of spatially rotated chiral spiral unit cells arranged with supercell periodicity N=3, engineered to produce controlled circular retardance. By rotating the effective coordinate frame on the Poincaré sphere, we formulate the transformation into a linear-eigenmode basis and show that spatial rotation of the spiral induces a geometric phase ramp analogous to a blazed grating. Crucially, this phase modulation persists for linear polarization inputs, revealing the basis independence of geometric phase accumulation.

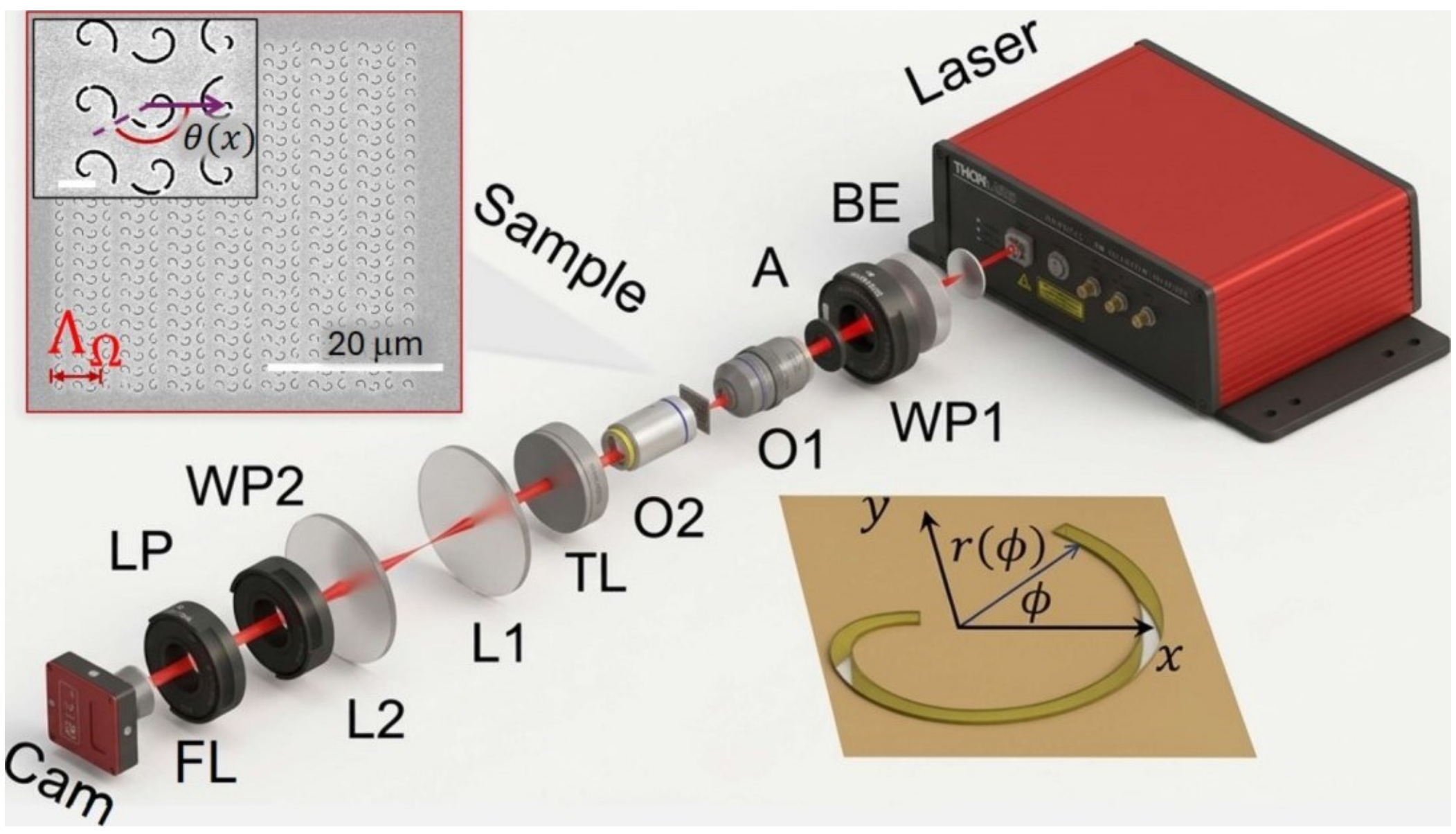

Figure I: Optical setup for real- and k-space imaging. The beam was expanded using a beam expander (BE) and then passed through the LRM system comprising of illumination objective (01) and the imaging oil immersion objective (02) attached to the substrate of the sample with an index matching oil. The imaging system included a tube lens (TL) a couple of relay lenses (Ll,L2) and a camera (Cam). An aperture (A) was used to control the beam size and the wave plate (WPI) controlled the input polarization. The Stokes parameters were determined using the classical four-measurement scheme with a quarter-wave plate (WP 2) followed by a linear polarizer (LP). A Fourier lens (FL) was used to image the k-space Insets show the scanning electron microscope image of the spiral array fabricated by the FIB and the schematic geometry of a single spiral slit.

A leakage radiation microscopy (LRM) [29, 30] followed by a Fourier lens is used to map the in-plane SP momentum distribution in the far field and directly resolve the geometric diffraction orders in k-space. These orders intensities are found to be strongly dependent on the rotation of the incident linear polarization. By measuring the local Stokes parameters of the emerged field, we confirm a space-variant field ellipticity induced by the structure. This effect exemplifies the basis transformation of the Poincaré space.

In general, our results establish a circular-birefringent analogue of PB metasurfaces and demonstrate that geometric phase modulation does not require linear birefringence nor circular polarization excitation. Our findings reveal a deeper complementarity between polarization basis and geometric phase accumulation, expanding the design space of metasurface-based phase engineering and opening opportunities for polarization-sensitive platforms in chiroptical spectroscopy and biophysical sensing. More broadly, our findings clarify the fundamental equivalence between spin-dependent and basis-independent geometric phase accumulation in structured chiral media.

# Observation of the Geometric Phase

We start by fabricating spiral slit arrays on a 60 nm thick gold film deposited on a glass substrate by sputtering (see inset in the Fig. 1). The structures were etched through the gold using a focused ion beam (FIB) technique. Each unit-cell was represented by a slit of an Archimedean spiral geometry with a periodicity of $\Lambda = 3825$ nm [31]. The spirals in the arrays were rotated along x: - $\theta(x) = \pi x/N\Lambda$, such that a $2\pi$ revolution was achieved within three periods (N=3). The radius of the spirals was given by $r(\varphi) = r0 + m\varphi/2\pi$, where m defined the winding number (pitch). We have tested arrays with m = 0.25, 0.5, 1 and the observed effects were consistent. Hereafter, we present the results for m=0.5 due to the fabrication convenience. We illuminated the array with a laser beam at 785 nm. The optical characterization was performed using the LRM system based on the immersive microscope objective attached to the glass substrate of the sample by means of an index matching oil. This system enabled us to study the plasmonic signal distribution in the real-space and the k-space (Fourier-space) by adding a Fourier Lens before the camera (see Fig. 1). The polarization of the incident beam was controlled using a half-wave plate (HWP) or a quarter-wave plate (QWP) rotated from $0^0$ to $355^0$ in steps of $5^0$ (72 angular positions). k-space intensity distributions were recorded for each input polarization state. A typical image (shown in Fig. 2) represents a characteristic diffraction pattern due to the periodic spiral array. The red crosses mark the ordinary diffraction orders, obtained at $(k\alpha, k\beta)$ — $(\alpha 2\pi/\Lambda, \beta 2\pi/\Lambda)$ positions (a, ß are integers). The dashed line emphasizes the plasmonic momentum $k_{SP} = k_0\sqrt{\frac{\epsilon_m}{\epsilon_m+1}}$ where $k0 = 2\pi/\lambda0$ is the laser beam wavenumber and $\epsilon_m = -22.88$ [32] is the real part of the dielectric constant of gold. Note, that each ordinary diffraction spot is surrounded by a couple of additional orders (hereafter we refer to them as satellites). The investigation of the satellites intensities versus wave plate rotation shows that these diffraction orders are polarization dependent. The positions along axis of the satellite spots are given $k_{SP} = k\alpha \pm 2\pi/N\Lambda$ which clearly confirms that they originate from a geometric phase [19].

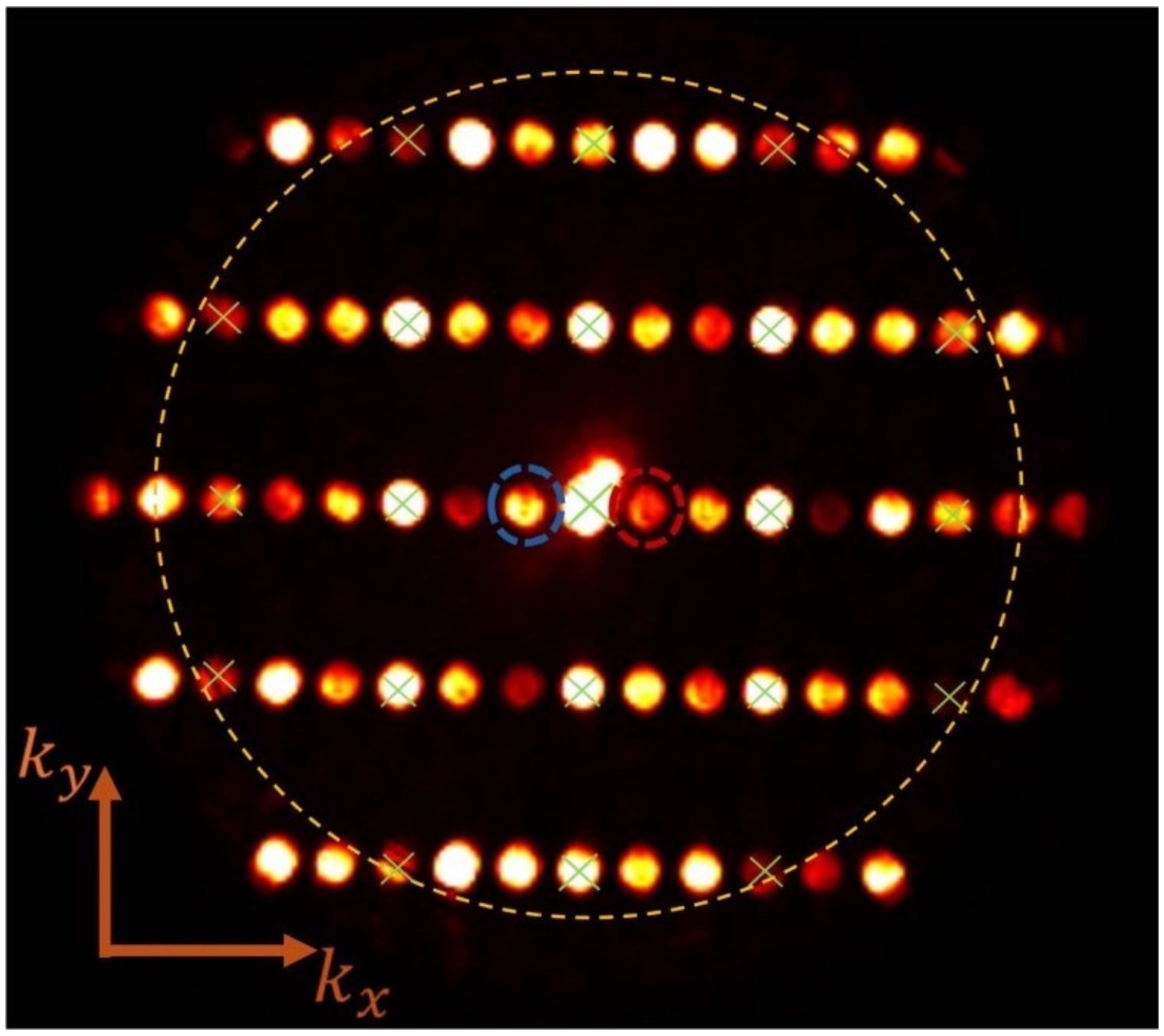

Figure 2: Typical image obtained in the k-space with a space variant grating. The dashed yellow circle depicts the plasmonic resonance. The blue and the red dashed circles indicate the satellite orders from which the intensity was measured. The cross marks correspond to the ordinary diffractions.

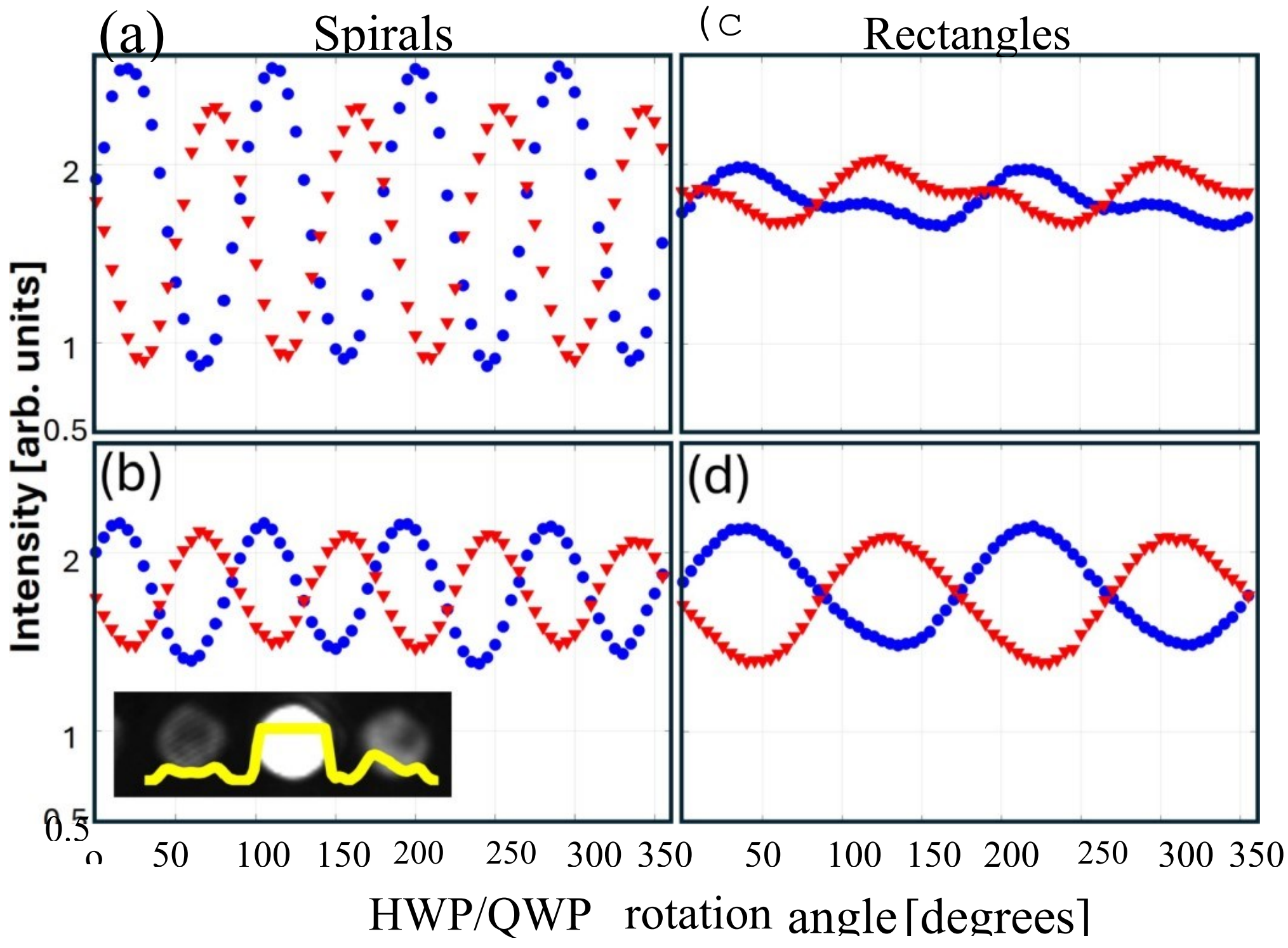

Figure 3: Polarization dependence of the satellite orders of the rotated spiral (a,b) & rectangular (c,d) structure measured by varying the orientation of (a) the HWP and (b) the QWP. The intensity of the different satellite orders are depicted by the inverted red triangles and blue circles. Inset shows intensity distributions in the k-space of the chosen orders at two angles of interest.

We are now interested in the particular mechanism responsible for these geometric orders. We plot the intensities measured at the satellite spots around the zero-order diffraction (marked by the blue and the red dashed circles) versus the waveplate angle. In Fig. 3 we compare the results in arbitrary units and in the same scale for rotating HWP (a) and QWP (b). A typical satellite cross-section is shown in the inset. The "anti-phase" behaviour is prominent in both cases, while the amplitude of beating is at least two times stronger in the HWP case. This observation emphasizes the dependence of the geometric phase induced by the spiral array on the orientation of the linear polarization. Similar effect was previously observed behind arrays of spatially rotated linear birefringence [7] or spatially rotated plasmonic slits [33].

We fabricated a sample consisting of the rotated rectangular aperture with the dimensons of 450 nm $\times$200 nm, while the period and the rotation rate were set to be similar to the spiral array. This arrangement resulted in a similar k-space distribution. In Fig. 3 c, d we append the measured satellite order efficiencies versus HWP QWP angle. We note that here, a strong anti-phase intensity variation is notable for the QWP rotation while the HWP rotation weakly alters both intensities simultaneously. In other words, the geometric phase orders now show a strong dependence on the light helicity. This is a straight-forward result of the classic optical spin-orbit interaction as depicted using a Poincaré representation of the polarization interaction (Fig. 4a). When a left-handed circular polarization is incident upon the space-variant nanoaperture, the emerging beam polarization can be approximated as a spatially rotated linear state. This field propagates to the far-field and produces a right-handed circularly polarized diffraction order. This example of the spin-orbit interaction is represented in the Figure as a blue meridional line. Moving along this line defines the ellipticity angle $2\chi$ stemming from the linear birefringence (LB) of the optical medium. The geometric phase between different slits is given by the half of the area enclosed by two corresponding meridians distinguished by the azimuthal angle $2\psi$ (here $\psi = \theta$). Since the rotation is a linear function of x a linear saw-tooth type phase arises behind the plasmonic structure resulting in the "satellite" orders. By altering the QWP angle the input polarization state can be varied from the right-hand to the left hand leading to the satellite orders to exchange intensity as shown in Fig. 3d. In the case of the spiral array the satellite order intensities are predominantly affected by the rotation of the linear state with extremum values about $\pm 45\,^{0}$ (here $|\nearrow\rangle$ and $|\searrow\rangle$). To analyze the origin of the measured geometric phase we propose a transformation of the Poincaré space as shown in Fig.4b. The sphere is rotated around the $S_1$ axis so that the north and the south poles represent $|\nearrow\rangle$, $|\searrow\rangle$ states. Accordingly, the interaction of these eigen-states with the structure could be mapped by the meridian yellow lines. This means that each spiral element produces a circular birefringence

(CB) whose rotation results in the ellipticity variation.

# Poincaré space transformation

Mathematically, we propose to treat the effect of the geometric phase induced by a space-variant circular birefringence using Poincareé space SO(3) rotation. We start with the standard right-handed Stokes basis:

$$S = \begin{bmatrix} S_1 \\ S_2 \\ S_3 \end{bmatrix} = S_0 \times \begin{bmatrix} cos2\psi cos2\chi \\ sin2\psi cos2\chi \\ sin2\chi \end{bmatrix}$$

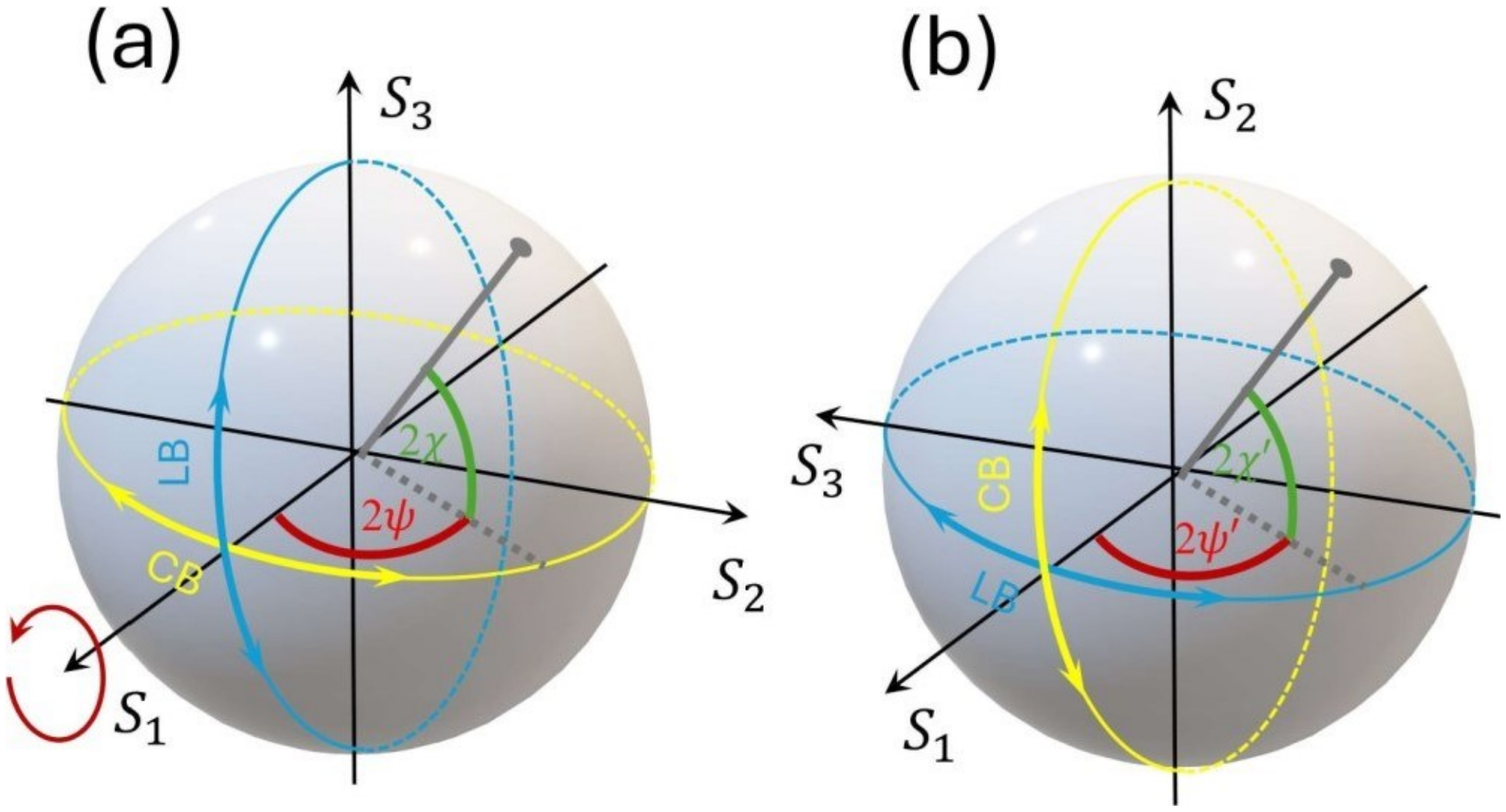

Figure 4: Poincaré sphere representation of the geometric phase resulted from a spatially rotated LB (a) and CB (b). The proposed Stokes parameters transformation map the angles $2\psi$, $2\chi$ into $2\psi'$ and $\chi'$.

where, $2\psi$ is a longitudinal (azimuthal) angle and $2\chi$ is a latitude (ellipticity) angle. The rotation about the $S_1$ axis by the angle $90^0$ is derived by means of the rotation matrix,

$$S' = \begin{bmatrix} S_1' \\ S_2' \\ S_3' \end{bmatrix} = \mathbf{R}(90^0)S = \begin{bmatrix} 1 & 0 & 0 \\ 0 & 0 & 1 \\ 0 & -1 & 0 \end{bmatrix} \begin{bmatrix} S_1 \\ S_2 \\ S_3 \end{bmatrix} = \begin{bmatrix} S_1 \\ S_3 \\ -S_2 \end{bmatrix}$$

maps the Stokes vector to:

$$\begin{aligned} S_1' &= S_1 = S_0 cos2\psi cos2\chi \\ S_2' &= S_3 = S_0 sin2\chi \\ S_3' &= -S_2 = -S_0 sin2\psi cos2\chi \end{aligned} \tag{2}$$

With these new definitions in mind we wish to redefine the new longitude and latitude angles in accordance with Eq. ,

$$\begin{aligned} \tan 2\psi' &= \frac{sin2\chi}{\cos 2\psi \cos 2\chi} \\ \sin 2\chi' &= \sin 2\psi \cos 2\chi \end{aligned} \tag{3}$$

These angles are depicted in Fig. 4b. In particular, the CB (retardance between the circular states) is now represented by the $\chi'$ while the LB (retardance between the linear states) corresponds to the angle $\psi'$.

When our system is illuminated by on eigen state polarization (for instance |↗⟩ the opposite eigen state ( |↘⟩ emerges with a geometric phase equal to the half of the area enclosed by the meridian lines. Therefore the ability of the plasmonic spiral array to manipulate the phase gradients depends on the variation of the LB along it. Here we propose to measure the ellipticity variation by a direct polarization measurement. Using the mapping introduced in this chapter we expect the geometric phase variation to be $\Phi g = \zeta 2\psi'$ with the pseudo-spin number $\zeta$ obtaining a value of +1 for the |↗⟩ and-1 for the |↘⟩ state. The laser beam

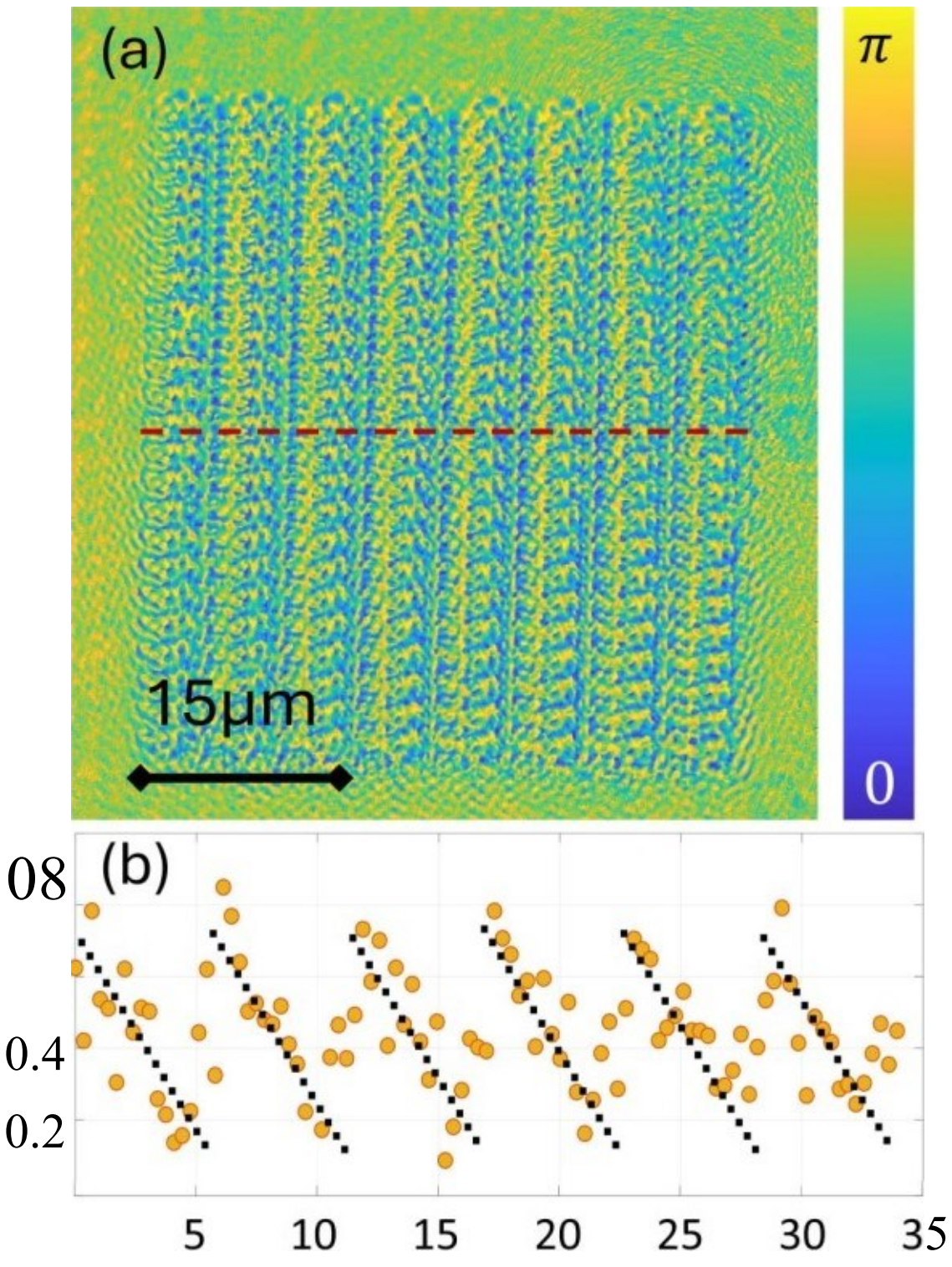

Figure 5: Polarization analysis behind the spirals array. An angle $\psi'$ distribution over the area of the sample is shown in (a) while its cross section along the red line is demonstrated in (b) along with the theoretically calculated geometric phase (dotted black line).

was expanded to fully cover the nanostructure area. The polarization state of the transmitted beam in the real space was reconstructed using the classical four-measurement Stokes technique 134, 35]. The first three Stokes parameters ($S_0$, $S_1$, $S_2$) were measured with the linear polarizer only, rotated to $\theta = 0^{\circ}$, $45^0$ , and $90^0$ .The fourth Stokes parameter ($S_3$) was obtained using a quarter-wave plate at 90 $^0$ followed by a polarizer oriented at $45^0$ .

The measured Stokes parameters were first normalized to So and the azimuthal angle of the transformed Poincaré state was obtained using the relations in Eq.3 . Figure 5a demonstrates the $\psi'$ map directly derived from the Stokes measurements. A periodic angle variation is clearly observed across the sample, while outside the structure its value b comes close to zero. We also plot the cross-section of the map across the dashed red line in Fig. 5b. The values seem to follow a cyclic saw-tooth behavior within the range of 0.8 $\pi$. A fundamental theory of a "blazed" grating suggests that reducing the effective phase gradient from $2\pi$ affects the diffraction efficiency of the 1st order. To account for the actual diffraction efficiency in our experiment we employ the visibility calculation for the data series obtained in Fig. 3a using

$$V_{max} = \left[ \frac{I_{+1} - I_{-1}}{I_{+1} + I_{-1}} \right]. \quad (4)$$

where are the measured intensities at the first-order satellites. We use a rigorous Matlab-based calculation to retrieve the theoretical phase that corresponds to the obtained orders visibility and plot it on top of the measured data in Fig. 5b. The results show a perfect correspondence to the theoretical geometric phase, which confirms the nature of the effect. Finally, we demonstrate an example of a plasmonic device based on the space-variant chirality. We fabricate a rotating spiral array with a period of 1020 nm and with N= 3. These parameters were chosen such that the satellites of the first diffraction order overlap with the plasmonic momentum, $k_{SP} = 2\pi\,\Lambda + 2\pi\,N\Lambda$ The inspection of the k-space (see Fig. 6a, b) reveals that a left-hand right-hand satellite maximize for state and provide a perfect coupling to a plasmonic unidirectional beam. This surface wave can be clearly seen in Fig. 6c, d. The yellow arrow marks the relevant geometric orders in the k-space and the resulting plasmonic beams in the real space. The white dashed line depicts the plasmonic momentum at our wavelength. The cross-section of the measured k-space along the central line is also shown, and the satellites intensity variation can be clearly observed. Similar device has been previously demonstrated with rotating rectangular plasmonic nanoapertures under circular state illumination [33]. Nevertheless, in the current system a space-variant chiral polarizer leads to the geometric phase from the linear states.

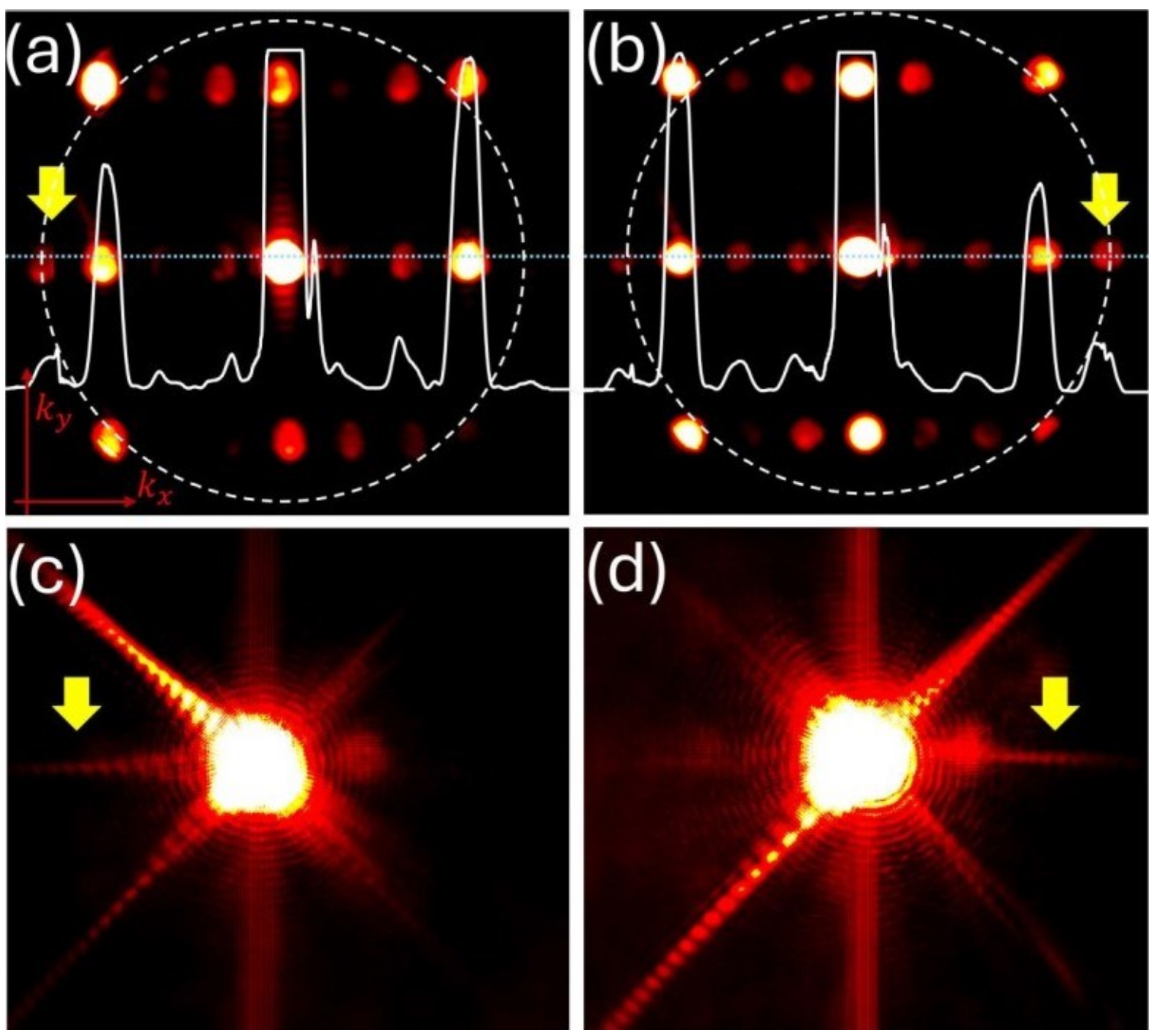

Figure 6: Plasmonic distribution measured from a rotating spiral array in the real space - (a,b) and in the k-space - (c,d) with and input states, respectively. The yellow arrow guides the eye to the dominant satellite/plasmonic beam. The white dashed line shows the SP momentum and the k-space cross-section taken along the blue dashed line is shown in the inset.

# Conclusions

In this work, we have experimentally demonstrated geometric phase modulation in plasmonic spiral slit arrays arising from circular birefringence and observed under linear polarization excitation. Using leakage radiation microscopy, we directly resolved the resulting diffraction satellites in Fourier space and established their polarization-dependent behavior.

One of the results of this study is the direct correspondence between the spatial polarization phase distribution and the momentum-space asymmetry of the diffracted field. By reconstructing the Stokes parameters and introducing a transformed Poincaré representation, we showed that the polarization evolution can be interpreted as an effective rotation around the $S_1$ axis. Within this framework, the modified phase parameter $\psi'$ exhibits a sawtooth-like spatial profile consistent with a geometric phase ramp.

Importantly, we demonstrate that the slope of this phase profile is quantitatively linked to the normalized intensity contrast between the $\pm1$ diffraction orders. This establishes a direct bridge between real-space phase gradients and Fourier-space energy redistribution, providing an experimentally accessible signature of geometric phase accumulation.

These results confirm that geometric phase modulation in chiral plasmonic systems is fundamentally basis-independent and can be fully described through polarization evolution on the Poincaré sphere. The demonstrated control of diffraction asymmetry and directional plasmon excitation highlights the potential of space-variant chiral structures for phase-engineered nanophotonic devices and polarization-sensitive applications.

# Acknowledgements

The authors thank the Israeli Ministry of Science Innovation and Technology and the Israeli Ministry of Alyiah and Integration for financial support.

## Funding

We acknowledge the Israeli Ministry of Science Innovation and Technology and the Israeli Ministry of Alyiah and Integration for the financial support.

## Disclosures

The authors declare no conflicts of interest.

## Data Availability

Data underlying the results presented in this paper are not publicly available at this time but may be obtained from the authors upon request.

## Supporting Information

No supporting information files are listed in this manuscript.